\documentclass[preprint,aps,amsmath]{revtex4}
\usepackage{graphicx}
\usepackage{bm}

\begin{document}
\preprint{}
\title{Drift wave turbulence in the presence of a dust density gradient}
\author{A. Kendl}
\affiliation{Institute for Ion Physics and Applied Physics, 
University of Innsbruck, A-6020 Innsbruck, Austria}
\author{P. K. Shukla}
\affiliation{RUB International Chair, International Centre for Advanced
  Studies in Physical Sciences, Faculty of Physics \& Astronomy, Ruhr
  University Bochum, D-44780 Bochum, Germany, and Department of Mechanical and Aerospace 
Engineering, University of California San Diego, La Jolla, CA 92093, USA \vspace{1cm}}

\begin{abstract}
\vspace{0.5cm} We present turbulent properties of electrostatic drift waves in a nonuniform
collisional plasma composed of magnetised electrons and ions in the presence
of immobile dust particles. For this purpose, we derive a pair of nonlinear
quasi-two-dimensional equations exhibiting the coupling between the
generalized  ion vorticity and  density fluctuations associated with
collisional drift waves.
The effect of a dust density gradient on the initial drift instability and fully
developed turbulence is examined numerically.

\vspace{6cm}

{\sl This is the preprint version of a manuscript submitted to Physical Review E (2011).}
\end{abstract}
\maketitle

\section{Introduction}

Low-frequency (in comparison with the ion gyrofrequency) electrostatic drift
waves in a nonuniform collisionless magnetized electron-ion plasma are
supported by inertialess electrons and inertial ions \cite{Kadomtsev65}.
In the pseudo-three dimensional drift wave electric field ${\bf E} =-\nabla \phi$, 
where $\phi$ is the wave potential, the electrons have a helical trajectory due 
to their ${\bf v}_E = (c/B_0^2){\bf E} \times {\bf  B}_0$ and ${\bf v}_D 
=-(ck_B T_e/eB_0^2 n_e )  {\bf B}_0 \times \nabla n_e$
drift motions in a plane perpendicular to the external magnetic field 
${\bf  B}_0 =\hat {\bf z} B_0$, as well as  their rapid motion along the external
magnetic field direction, where $\hat {\bf z}$ is the unit vector along the
$z$ axis in a Cartesian coordinate system, $B_0$ the strength of the magnetic
field, $c$ the speed of light in vacuum, $k_B$ the Boltzmann constant, $e$ the
magnitude of the electron charge, $T_e$ the electron temperature, and $n_e$
the electron number density.

\bigskip

Furthermore, the motion of the two-dimensional
ions in a plane perpendicular to the axial direction is governed by the ${\bf
  v}_E$ and the polarization ${\bf v}_p =-(c/B_0 \omega_{ci}) d_t \nabla_{\perp}^2
\phi$ drifts, where we have denoted $d_t = \partial_t + {\bf
  v}_E \cdot \nabla$, $\omega_{ci} =eB_0/m_ic$ is the ion gyrofrequency, and
$m_i$ the ion mass. 

\bigskip

Hence, the electron and ion density fluctuations are
different due  to the differential electron and ion motions in a magnetized
plasma with an equilibrium density gradient. The resulting space charge
separation, in turn, causes  dispersive drift oscillations to propagate across
the homogenous magnetic field and density gradient directions.

\bigskip

In a collisional electron-ion magnetoplasma, dispersive drift waves are excited by
the combined action of the density inhomogeneity and electron-ion
collisions. The latter produce a phase lag between the magnetic field-aligned
electron velocity perturbation as well as the drift wave potential and the
electron density perturbation. As a result, drift waves grow exponentially,
extracting energy from the equilibrium density gradient. 

\bigskip

It is understood that
nonthermal drift waves cause anomalous cross-field diffusion of the plasma
particles \cite{Kadomtsev65,Horton99,Weiland00}. The plasma particle
confinement is significantly improved if there are cylindrically symmetric
zonal/sheared flows \cite{Shukla81,Shukla02a,Diamond05} that are nonlinearly excited by
finite amplitude dispersive drift waves in plasmas. Thus, zonal flows act like
a transport barrier \cite{Tynan09}.

\bigskip

Charged dust impurities are common in space and laboratory plasmas
\cite{Shukla02b,Shukla09,Krash11}. The presence of charged dust grains in an
electron-ion plasma modifies the equilibrium quasi-neutrality condition
$n_{i0} =n_{e0} +\epsilon Z_d n_{d0}$, where $n_{j0}$ is the unperturbed
number density of   the particle species $j$ ($j$ equals $i$ for ions, $e$ for 
electrons, and $d$ for charged dust grains), $Z_d$ is the number of constant 
charges residing on dust, and
$\epsilon = 1$ $ (-1)$ for negative (positive) dust. Due to the modification
of the quasi-neutrality condition, the difference between the divergence of
the  electron and ion fluxes involving the ${\bf V}_E$ drift is finite. This
leads to a flute-like Shukla-Varma (SV) mode \cite{Shukla93} in a nonuniform dusty
magnetoplasma. The SV mode governs the dynamics of the ion
vorticity, which evolves in the form of a dipolar vortex.

\bigskip

In this paper, we present an investigation of the pseudo-three dimensional
dissipative drift wave turbulence in a nonuniform collisional dusty
magnetoplasma  composed of the electrons, ions and stationary charged dust
grains. By using the two-fluid model and the guiding center drift
approximation (viz. $|d/dt| \ll \omega_{ci}$), we derive a pair of nonlinear
generalized Hasegawa-Wakatani (HW) equations \cite{Hasegawa83}
governing the evolution of the potential and density fluctuations
associated with the  low-frequency (in comparison with $\omega_{ci}$), long
wave length (in comparision with the thermal ion gyroradius) collisional drift
waves in our dusty plasma. 

\bigskip

Generalized HW type equations in the presence of dust have before
been derived and linearly analyzed in refs.~\cite{Benkadda96} and \cite{Vladimirov98}.
A linear analysis of dissipative drift modes in the presence of a dust density
gradient has been presented in Ref.~\cite{Benkadda02}, and further discussed in
Ref.~\cite{Manz10} with respect to nonmodal growth, including dust charging effects.

\bigskip

The work presented in refs.~\cite{Benkadda96,Vladimirov98,Benkadda02,Manz10}
(and others for similar topics) has shown that a dynamically variable dust
charge can have a profound effect on linear drift wave stability.
In a comprehensive dusty plasma turbulence model the charging process should
therefore be taken into account.
We have here refrained from including the charging process in order not
to multiply the number of free parameters beyond the absolute necessary.

\bigskip

The effect of dust charging dynamics on (nonlinear fully developed)
drift wave turbulence has (to our knowledge) not yet been addressed in the
literature. While we recognize that this would indeed be a worthwile effort,
it is another kind of study itself 

\bigskip

An important point why we have so far refrained from performing this study
ourselves is the many uncertainties related to the dust grain charging model 
for a nonuniform, collisional magnetoplasma that have been used so far. In particular,
the dust grain charging is a dyanmical process and requires a complete
kinetic theory for fluctuating (in the electrostatic field of pseudo-three-dimensional 
drift waves) electron and ion currents that reach the surface of dust grains in our
plasma.


\bigskip

In the following, we study turbulence properties of the modified (by the
presence of charged dust particles) drift waves in a nonuniform collisional
magnetoplasma with static charged dust impurities. Thus, we are assuming that
the timescales for the excitation of drift waves and their evolution are much
shorter than the dust plasma and dust gyroperiods, and much longer than the
electron-ion relaxation period. 

\bigskip

Furthermore, the assumption of constant dust charge would
remain valid since the dust grain charging frequency is typically much larger
than the modified drift wave frequency. Thus, our nonlinear simulation study
of the dissipative drift wave turbulence is based on a slightly different set
of dusty HW-SV equations compared to
refs.~\cite{Benkadda96,Vladimirov98,Benkadda02,Manz10}, excluding dust charge
fluctation effects.  
Our simulation results reveal that the equilibrium dust density gradient controls
the formation of turbulent structures and the associated cross-field transport
of the plasma particles in a nonuniform dusty magnetoplasma. 

\bigskip

The manuscript is organized in the following fashion. In Sec. II, we derive
the governing nonlinear equations for the low-frequency (in comparision with
the ion gyrofrequency) dissipative drift wave turbulence in a nonuniform dusty
magnetoplasma composed of magnetized electrons and ions, as well as
unmagnetized stationary charged dust grains. We use the 
two-fluid equations in the guiding center approximation to derive a pair of
nonlinear equations 
that exhibits coupling between the plasma density perturbation and the drift
wave potential. Section III presents a local linear dispersion relation and
its analysis. Governing equations for nonlinearly interacting finite amplitude
drift waves are numerically analyzed in Sec. IV. 
Summary and conclusions are contained in Sec. V.

\section{Nonlinear drift wave equations}

Let us consider the electrostatic drift waves with the electric field 
${\bf E} =-\nabla_{\perp} \phi -\partial_z \phi$ in a plasma with a
uniform magnetic field $\hat {\bf z} B_0$ and the equilibrium  density
gradient $\partial_x n_{j0}$. 

\bigskip

In a low-beta ($\beta=8\pi n_{e0}k_B
T_e/B_0^2 \ll 1$)  plasma, the electron and ion fluid velocities for  
$|d/dt| \ll \omega_{ci}, \nu_{e}$ are, respectively,

\begin{equation}
{\bf u}_e \approx \frac{c}{B_0} \hat {\bf z} \times \nabla \phi- \frac{ck_B
  T_e}{eB_0 n_e} \hat {\bf z} \times \nabla n_e + \hat {\bf z} u_{ez},
\label{e-1}
\end{equation}

and

\begin{equation}
{\bf u}_i \approx \frac{c}{B_0} \hat {\bf z} \times \nabla \phi -\frac{c} {B_0
  \omega_{ci}} \left( \frac{d}{dt}  + \nu_i\right) \nabla_{\perp} \phi,
\label{e-2}
\end{equation}

with the magnetic field-aligned electron fluid velocity perturbation is given by

\begin{equation}
u_{ez} =\frac{e}{m_e\nu_e}\frac{\partial}{\partial z}\left( \phi - \frac{k_B
  T_e}{e} \frac{n_{e1}}{n_{e0}}\right), 
\label{e-3}
\end{equation}
where $m_e $ is the electron mass, $\nu_e$ ($\ll \omega_{ce} =eB_0/m_e c$) the
electron-ion collision frequency, $\nu_i$ the ion-neutral collision frequency,
and $n_{e1} =n_e - n_{e0}$ ($\ll n_{e0}$) is a small perturbation in the
electron number density. 

\bigskip

The two-dimensional (2-D) ions are assumed to be
cold. The assumption of 2-D ions ensures the decoupling of the dust
ion-acoustic and drift waves 

\bigskip

Inserting eqs.~\ref{e-1} - \ref{e-3} into the electron and ion continuity equations,
we obtain the evolution equations for the electron and ion number density
perturbations $n_{e1}$ ($\ll n_{e0}$) and $n_{i1}$ ($\ll n_{i0}$), respectively 

\begin{equation}
\frac{d n_{e1}}{dt} 
-\frac{c}{B_0} \hat {\bf z} \times \nabla n_{e0} \cdot \nabla \phi 
+ \frac{n_{e0}e}{m_e \nu_e} \frac{\partial^2}{\partial z^2} 
\left( \phi - \frac{k_B T_e}{e} \frac{n_{e1}}{n_{e0}} \right), 
\label{e-4}
\end{equation}

and

\begin{equation}
\frac{d n_{i1}}{dt} 
-\frac{c}{B_0} \hat {\bf z} \times \nabla n_{i0} \cdot \nabla \phi 
- \frac{c n_{i0}e}{B_0 \omega_{ci}} 
\left( \frac{d}{dt} + \nu_i\right ) \nabla_{\perp}^2 \phi.
\label{e-5}
\end{equation}

\bigskip

Now, subtracting eq.~\ref{e-5} from eq.~\ref{e-4} we obtain the generalized
ion vorticity equation

\begin{equation}
\left( \frac{d}{dt} + \nu_i \right) \nabla_{\perp}^2 \phi 
+ \omega_{ci} K_d \frac{\partial \phi}{\partial y} 
+\frac{\Omega_{LH}^2}{\nu_e} \frac{\partial^2}{\partial z^2}
\left( \phi - \frac{k_B T_e }{e} \frac{n_{e1}}{n_{e0}} \right), 
\label{e-6}
\end{equation}
where $K_d =(\epsilon Z_d c/B_0n_{i0})\partial_x n_{d0}$, 
$\Omega_{LH} =\left(\omega_{ce} \omega_{ci}/\alpha \right)^{1/2}$ is the
lower-hybrid resonance frequency in our dusty plasma, 
the drift scale is $\rho_s =c_s/\omega_{ci}$, and $c_s = (\alpha k_B
T_e/m_i)^{1/2}$ is the modified ion-sound speed \cite{Shukla92}, with
$\alpha =n_{i0}/n_{e0} > 1$.
\bigskip

We note that the $K_d$ term arises due to a non-zero value coming from the difference 
between the divergence of the electron and ion fluxes involving the $(c/B_0){\bf E} \times
\hat {\bf z}$  drift in our nonuniform dusty plasma 
In the absence of the magnetic field-aligned electron motion,
Eq.~\ref{e-6} describes the dynamics of the SV mode \cite{Shukla93}.
In a quasi-neutral dusty plasma with immobile charged dust grains,
eqs.~\ref{e-4} and \ref{e-6} are closed with the help of $n_{i1}=n_{e1} \equiv
n_1$, which is valid as long as the frequency (wavelength) of the drift waves
is much larger than the plasma and dust gyrofrequencies (the electron Debye radius).
\bigskip

It is appropriate to normalize the wave potential $\phi$ by $k_B T_e/e$, the
density perturbation $n_1$  by $n_{e0}$, as well as the time and space
variables by the ion gyroperiod $\omega_{ci}^{-1}$ and the effective ion sound
gyroradius $\rho_s$, respectively. Then, the governing nonlinear equations for
the collisional drift waves in our nonuniform dusty magnetoplasma are

\begin{eqnarray}
\left( \frac{d}{dt} + \gamma \right) \Omega + a \frac{\partial \phi}{\partial
  y} - g (\phi -n) & = & 0, \label{e-7} \\ 
\frac{d n}{dt} + b \frac{\partial\phi}{\partial y}  - \left(\frac{d}{dt} +
  \gamma\right) \Omega & = & 0, 
 \label{e-8} 
\end{eqnarray}
where  $\Omega =\nabla_{\perp}^2 \phi$,  $n= n_1/n_{e0}$, $\gamma = \nu_i / \omega_{ci}$,
$a = K_d \rho_s $, $b = K_i \rho_s$, where $K_i =-(1/n_{e0})\partial n_{i0}/\partial x > 0$.
We have Fourier decomposed the drift wave  potential and electron density
perturbations along the $z$-axis, but kept their variations in the $x-y$
plane, so that $g = k_z^2 \omega_{ce}/ \nu_e \alpha$ is the dissipative
coupling coefficient. 

\bigskip

Equations (\ref{e-7}) and (\ref{e-8}) are the generalization of
the HW equations \cite{Hasegawa83}. The latter are recovered in the limit $a=0$ 
and $\gamma=0$.

\section{Linear dispersion relation}

Linearization of eqs.~(\ref{e-7}) and (\ref{e-8}) with an ansatz
$\phi, n \sim \exp[-i\omega t + i{\bf k}\cdot {\bf x}]$
obtains the dispersion relation

\begin{equation}
\omega^2 k^2 +  \omega A - \gamma g k^2 + i \omega G - i g B.
\end{equation}
where $A=a k_y$, $B=b k_y$ and $G=g k^2 + \gamma k^2 +g$.
Inserting $\omega = \omega_R + i\omega_I$ and solving the imaginary part
of the dispersion relation for $\omega_I \ll \omega_R$ results in the real frequency

\begin{equation}
\omega_R  = {gB \over G} = {b k_y \over 1 + (1+\gamma/g)k^2}.
\end{equation}

The growth rate is obtained from the real part as
\begin{equation}
\omega_{I} = {-1 \over K k} \left[ 1 \mp 
\sqrt{ 1 +  K^2 [ (\omega_R^2 + \gamma g) k^2 + A \omega_R ] } \right]
\end{equation}
with $K=2k/G$. Only the negative root is relevant for unstable solutions ($\omega_I >0$).

\bigskip

Instability thus occurs if
\begin{equation}
(\omega_R^2 + \gamma g) k^2 + A \omega_R   > 0.
\end{equation}

\bigskip

This is always achieved if $A \omega_R \sim AB \sim ab \geq 0$. The usual
resistive drift wave instabilty (expressed by the first term) is thus enhanced
for co-aligned gradients with $a>0$ if $b>0$.
\bigskip

The growth rate is reduced for counter-aligned gradients.
Drift modes are damped ($\omega_I < 0$) for a large counter-aligned dust
density gradient, when $|a| > [k/(1+k^2)]^2 |b|$. For $k=1$ this corresponds
to $a < - b/4$. 

\bigskip

As drift wave turbulence is most active around $k \approx 1$,
it can be expected that (for $b\equiv 1$) the turbulence is strongly damped for $a <
-0.25$, and enhanced approximately proportional to $\sqrt{a}$ for positive $a$ 

\bigskip

\begin{figure}
\includegraphics[height=6.0cm]{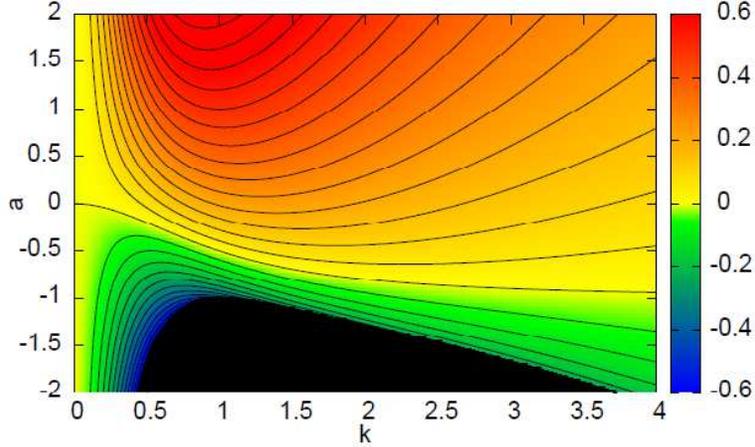}
\caption{\sl Growth rate $\omega_I(k,a)$ as a function
of wave number $k$ and dust density gradient $a$. Black marks the region of
stability. \vspace{1cm}}
\label{fig-lin}
\end{figure}

In Fig.~\ref{fig-lin} the growth rate $\omega_I(k,a)$ is plotted as a function
of wave number and dust density gradient.

\section{Nonlinear numerical solutions}

For numerical simulations, we write Eqs.~(\ref{e-7}) and (\ref{e-8}) as
\begin{eqnarray}
\partial_t \Omega + [ \phi, \Omega ] & = & -a \; \partial_y \phi  
                                    - \gamma \; \Omega + g \; ( \phi - n),
\label{dhw1} \\
\partial_t n + [ \phi, n ] & = & - (b+a) \; \partial_y \phi  + g \; ( \phi - n ),
 \label{dhw2} 
\end{eqnarray}
where the advection nonlinearity $[A,B] =\hat {\bf z} \times \nabla A \cdot \nabla B$
is expressed in terms of the Poisson brackets.
The equations are numerically solved with a third order
Karniadakis time stepping scheme in combination with the Arakawa method for
the Poisson brackets \cite{Karniadakis,Arakawa,Naulin03}.
Hyperviscuous operators $\nu^4\nabla^4$  with $\nu^4= -2 \cdot 10^{-4}$ are
added  for numerical stability to the right-hand side of both
equations \eqref{dhw1} and \eqref{dhw2}, acting on $\Omega$ and $n$,
respectively. The Poisson equation is solved spectrally.

The equations are for the turbulence computations discretized on a 
doubly periodic 512 $\times$ 512 grid with box dimension $L_x = L_y=64 \rho_s$.
The initial drift wave development are computed on a  128 $\times$ 256 grid
with box dimension $L_x = 64 \rho_s$ and  $L_y = 128 \rho_s$ 
Nominal plasma parameters are $g=0.5$ and $b=1$.

\section{Vortex development}

First, we study the initial evolution of a drift wave growing out of a density
perturbation in the presence of a dust density gradient $a$.

\bigskip

  The ion viscosity is
set to $\gamma = 0.1$. A Gaussian density perturbation is initialized with amplitude
$n=1$, which generates a potential perturbation of the same initally Gaussian form 
This leads to the formation of an ${\bf E}\times {\bf B}_0$ drift vortex, where the
plasma is convected around the perturbation.

\bigskip

\begin{figure}
\includegraphics[width=11.0cm]{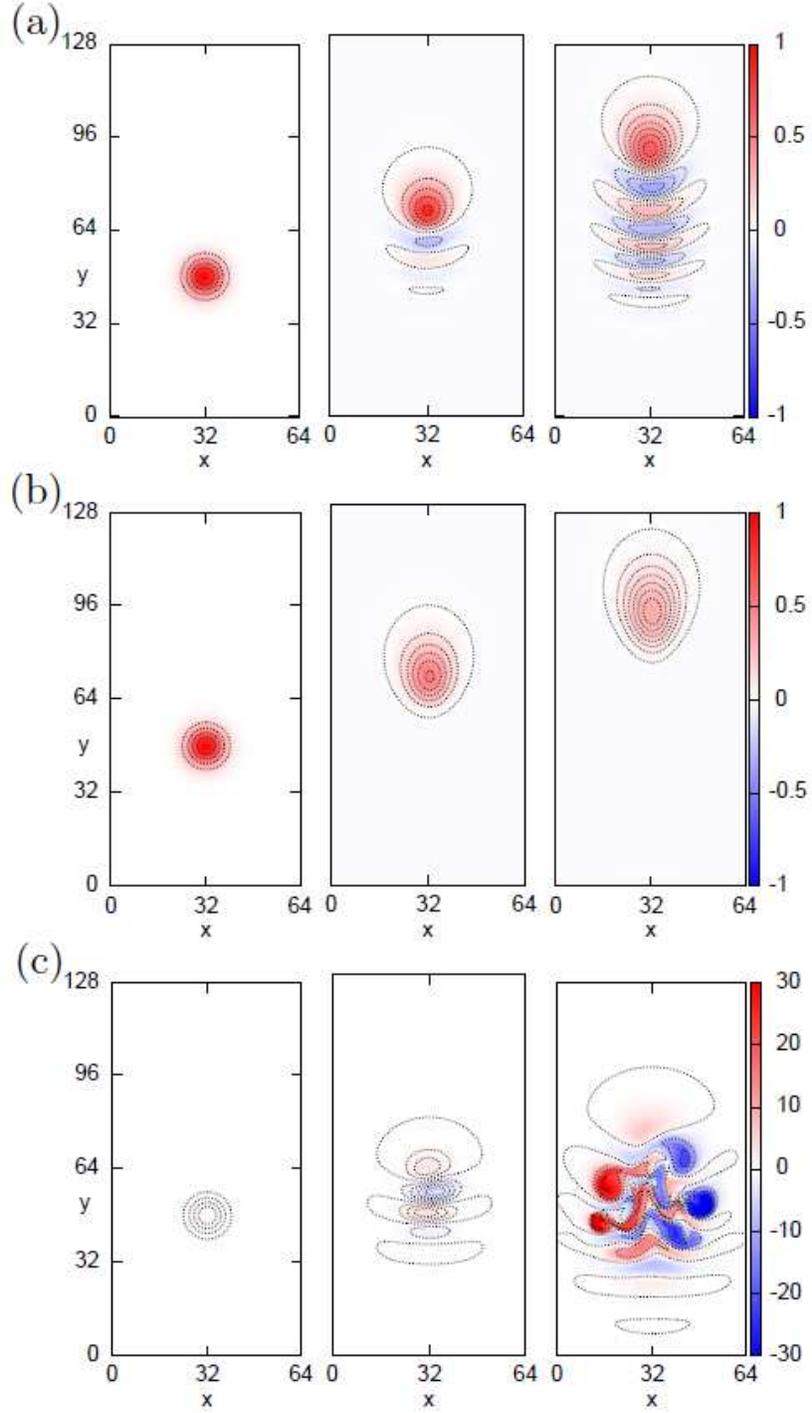}
\caption{\sl Development of a drift vortex with an  initial Gaussian density perturbation
 $n(x,y)$ in the presence of a dust density gradient $a$ for $t=0$ (left),
 $t=25$ (middle) and $t=50$ (right): (a) $a=0$, (b) $a=-0.5$, (c)
 $a=+0.5$. Note: the color scale (red for positive, blue for negative density
  perturbations in drift scaled units) for $n(x,y)$ in case (c) is 30 times larger
 than in (a) and (b).}
\label{fig-blobs}
\end{figure}

For $a=0$ the usual drift wave dynamics is encountered, which is shown in
Fig.~\ref{fig-blobs}a: the singular density structure ($t=0$, left frame) is propagating 
in the electron diamagnetic drift direction (upwards, in the positive $y$ direction)
and generating drift wave structures in its wake, which are shown for
times $t=25$ (middle frame) and $t=50$ (right frame). 

\bigskip

At a later stage, the drift wave nonlinearly develops via secondary
instabilities into a saturated turbulent state, whose properties are discussed
in the next section.

\bigskip

In Fig.~\ref{fig-blobs}b the evolution for $a=-0.5$ is shown: the vortex is
elongated in the $y$ direction, while the structure is propagating but damped.
The mode is still damped when the viscosity $\gamma$ is set to zero and the
initial amplitude to $n_0=10$.

\bigskip

Figure~\ref{fig-blobs}b shows the case of $a=+0.5$: The structure is much slower
propagating in the $y$ direction, but is forming a nearly standing drift wave structure,
which is elongated in the $x$ direction, and growing fast in amplitude until nonlinearly
secondary modes form by the Kelvin-Helmholtz break-up of the streamers.

\bigskip

The linear and nonlinear development of the drift wave perturbations thus strongly
depends on the sign of the dust density gradient. Co-alignment of the plasma
density gradient with the dust density gradient is enhancing the mode growth, while
counter-alignment is strongly damping.

\newpage
\section{Turbulent state}

The computations are started from a random quasi-turbulent spectral bath and
run into saturation. The spectral bath is generated initialising the density
field in Fourier space by (within some range) random amplitudes that approximately
follow a power law spectrum. In this way all $k$ modes are excited from start, and a
fully developed turbulent state is reached rather rapidly in the simulation.

\bigskip

Statistical averages are taken in the interval between
$200 \leq t \leq 1000$, which for all parameters is well in the saturated regime.

\bigskip

\begin{figure}
\includegraphics[width=10.5cm]{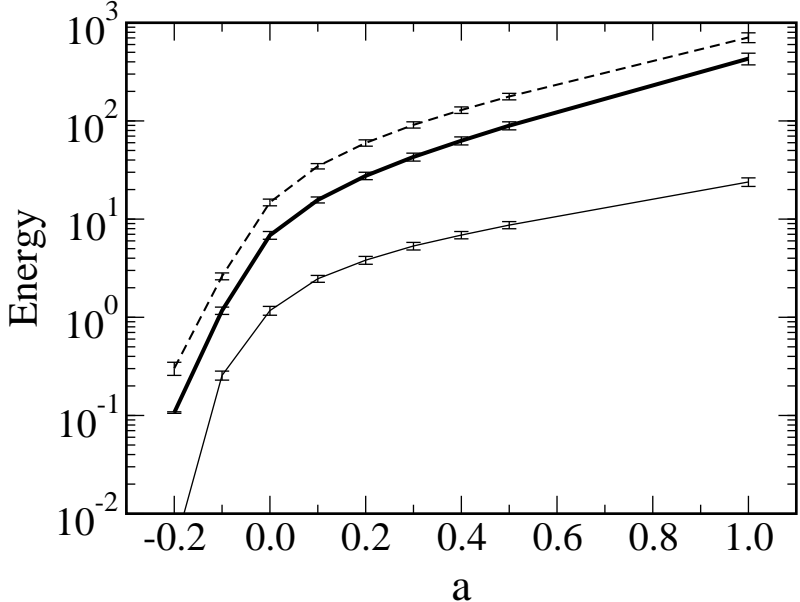}
\caption{\sl The potential energy $U$ (bold), the kinetic energy $K$ (thin) and
 the  generalized enstrophy $W$ (dashed) as a function of the dust density gradient $a$.}
\label{fig-energies}
\end{figure}

The turbulent activity can be characterised by the potential energy $U =
(1/2)\int dV n^2$, the kinetic energy $K = (1/2)\int  dV (\nabla \phi)^2$ and
the generalized enstrophy $W = (1/2) \int dV (n-\Omega)^2$.
The mean of these energetic quantities together with the standard deviation of
the fluctuations in the saturated state are displayed in Fig.~\ref{fig-energies}
for various values of the dust gradient parameter $a$. 

\begin{figure}
\includegraphics[width=10.5cm]{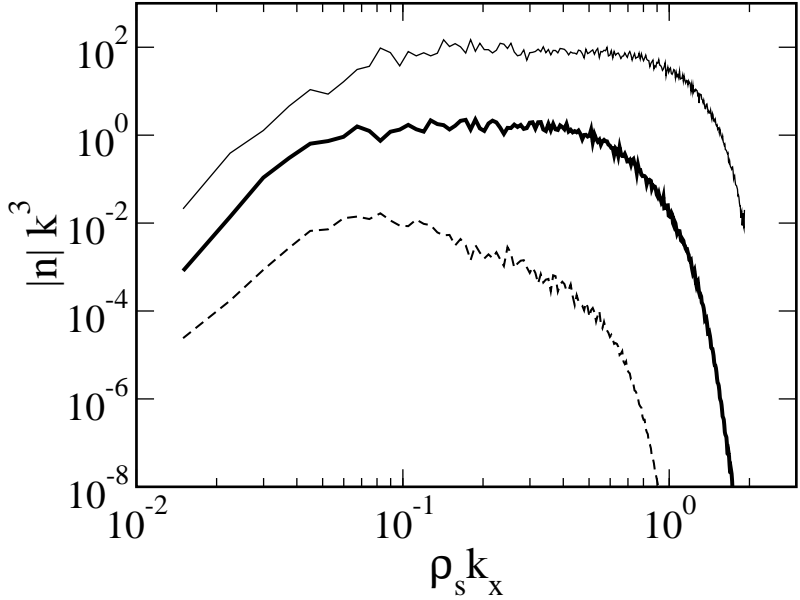}
\caption{\sl The wave number spectra of density fluctuations in the $x$ direction,
  multiplied by a factor $(k_x \rho_s )^3$: $a=0$ (bold line), $a=1$,
  $a=-0.2$ (dashed line).
}
\label{fig-spectra}
\end{figure}

\bigskip

The turbulent energies closely follow the predictions from the linear analysis: 
negative values of $a$ (counter-aligned gradients) have a strong damping
influence on the turbulent fluctuations. Co-aligned gradients are enhancing
the drive of the drift wave turbulence.

\bigskip

The wave number spectra $|n(\rho_s k_x)|(\rho_s k_x)^3$ are shown in
Fig.~\ref{fig-spectra}.  
The standard HW case ($a=0$, bold line) is arbitrarily scaled to 1.
For a co-aligned dust density gradient ($a=1$, thin line) modes around $0.5
\leq \rho_s k_x \leq 1$ in the order of a few drift scales $\rho_s$ are in
relation more strongly pronounced.
For counter-aligned gradients ($a=-0.2$, dashed line) the smaller scales are
most strongly damped.

\begin{figure}
\includegraphics[width=13.7cm]{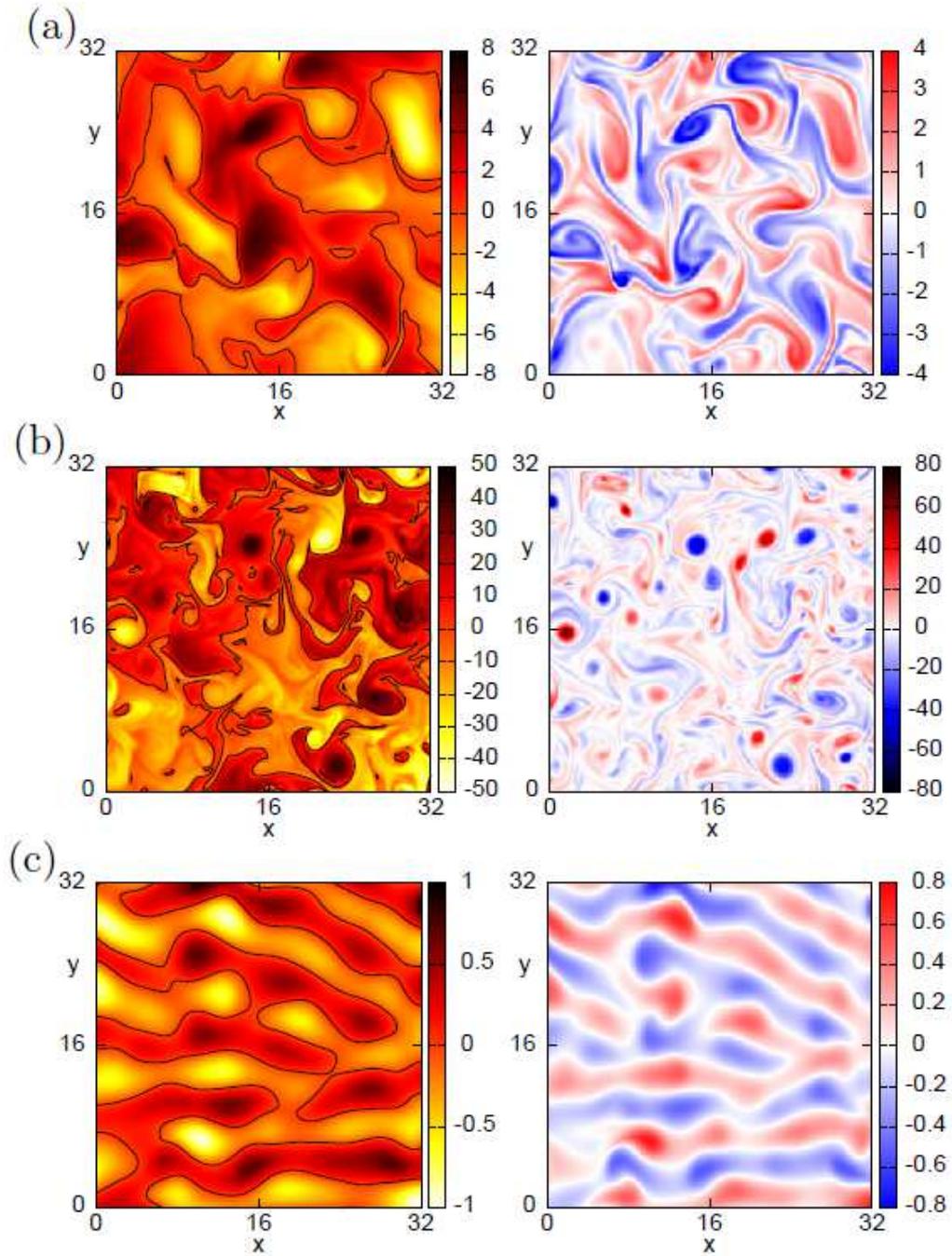}
\caption{\sl Density $n(x,y)$ (left) and vorticity $\Omega(x,y)$ (right) 
  for (a) $a=0$, (b) $a=1$, (c) $a=-0.2$. Only a quarter of the actual
  computational domain is shown here.\vspace{1cm}}
\label{fig-turb}
\end{figure}

\bigskip

This change in structural scales is also visible in 2-D $x-y$ plots of density
perturbations (left column) and vorticity (right column) in Fig.~\ref{fig-turb}.
The top row shows the standard case for $a=0$ of the typical drift wave turbulence.
In the middle row, a strong co-aligned dust density gradient $a=1$ leads to
a stronger perturbation of structures at smaller scales, visible by a more
frayed out density field (contour line drawn for $n=0$) and more strongly
pronounced small-scale vorticity structures. The bottom row shows the case of $a=-0.2$ 
with counter-aligned gradients, where quasi-linear drift wave structures appear,
which extend in the $x$-direction and propagate in the $y$-direction.

\newpage
\section{Summary and discussions}

Summarizing, we have investigated the properties of nonlinearly interacting
finite amplitude electrostatic drift waves in a nonuniform collisional dusty
magnetised plasma composed of 3-D magnetized  electrons, 2-D magnetized ions, 
and immobile charged dust impurities.
The dynamics of the dissipative drift wave turbulence in our model of a dusty
magnetised plasma is governed by generalized Hasegawa-Wakatani equations 
In the latter, one encounters nonlinear coupling between the quasi-neutral
electron density fluctuation and the ion vorticity that are driven by the
combined action of the dust density gradient and a dissipative electron
current arising from the magnetic field-aligned electron motion reinforced by
the parallel electric force and the parallel variation of the electron
pressure  perturbation for the constant electron temperature.

\medskip

Numerical simulations of the governing nonlinear equations reveal features of
the fully developed drift wave turbulence 
The presence of a co-aligned dust density gradient is found to enhance 
the drift wave turbulence and the cross-field plasma particle transport.
On the other hand, counter-aligned gradients lead to a damping of the drift
fluctuations. Thus, the dust density inhomogeneity plays a decisive role in
the formation of drift wave vortex structures in nonuniform dusty magnetised plasmas 

\medskip

In conclusion, we stress  that the  present results should be helpful  in
understanding the features of fully developed  low-frequency drift wave
turbulence that might  emerge from the forthcoming laboratory dusty plasma
experiments in an external magnetic field, and from nonuniform magnetized
space dusty plasmas.

\smallskip

\section*{Acknowledgements}

The work of AK was supported by the Austrian Science Fund (FWF) Y398
and by a junior research grant from University of Innsbruck.
The research of PKS at UCSD was supported by NSF PHY0903808.
The authors thank M. Rosenberg (Department of Electrical Engineering,
 University of California San Diego) for valuable discussions.

\end{document}